\documentclass[twocolumn,aps,prl,superscriptaddress,floatfix,longbibliography]{revtex4-1}  % this version shows reference title.
\usepackage[draft]{changes}
\usepackage[normalem]{ulem}
\usepackage{graphicx}% Include figure files
\usepackage{amssymb} % symbols
\usepackage{dcolumn}% Align table columns on decimal point
\usepackage{bm}% bold math
\usepackage[mathlines]{lineno}% Enable numbering of text and display math
\usepackage[colorlinks,linkcolor=blue,anchorcolor=blue,citecolor=blue,urlcolor=blue]{hyperref}
\usepackage{multirow}
\usepackage{color}
\usepackage{amsmath}

\makeatletter
\renewcommand*{\@fnsymbol}[1]{\ensuremath{\ifcase#1\or \dagger\or *\or \ddagger\or
\mathsection\or \mathparagraph\or \|\or **\or \dagger\dagger \or
\ddagger\ddagger \else\@ctrerr\fi}} \makeatother

\usepackage{color}%

\usepackage[draft]{changes}
\usepackage[normalem]{ulem}
\definechangesauthor[name={gao}, color=blue]{gao}

\usepackage{color}%

\begin{document}
\title{
%Unveiling Ultra-Low Thermal Conductivity and High Thermoelectric Performance in Tl$_9$SbTe$_6$ via Machine-Learned Potential\\
%{\color{blue} Machine learning for predicting ultralow thermal conductivity and high figure of merit in Tl$_{10-x}$Sb$_x$Te$_6$} \\
%{ Machine learning for predicting ultralow thermal conductivity and high \textit{ZT} in Tl$_9$SbTe$_6$ at medium temperatures} 
{Machine learning for predicting ultralow thermal conductivity and high \textit{ZT} in complex thermoelectric materials}
}

\author{Yuzhou Hao}
%\email[E-mail: ]{haoyuzhou@stu.xjtu.edu.cn}
\thanks{These authors contributed equally to this work.}
\affiliation{State Key Laboratory for Mechanical Behavior of Materials, School of Materials Science and Engineering,
Xi'an Jiaotong University, Xi'an 710049, China}
%\affiliation{These authors contributed equally to this work.}

\author{Yuting Zuo}
%\email[E-mail: ]{yt.zuo.2022@stu.xjtu.edu.cn}
\thanks{These authors contributed equally to this work.}
\affiliation{State Key Laboratory for Mechanical Behavior of Materials, School of Materials Science and Engineering,
Xi'an Jiaotong University, Xi'an 710049, China}
%\affiliation{These authors contributed equally to this work.}

\author{Jiongzhi Zheng} 
\thanks{These authors contributed equally to this work.}
\email[E-mail: ]{jiongzhi.zheng@dartmouth.edu}
\affiliation{Thayer School of Engineering, Dartmouth College, Hanover, New Hampshire, 03755, USA}
\affiliation{Department of Mechanical and Aerospace Engineering, The Hong Kong University of Science and Technology, Clear Water Bay, Kowloon, 999077, Hong Kong}
%\affiliation{These authors contributed equally to this work.}
%\email[E-mail: ]{jiongzhi.zheng@dartmouth.edu}

\author{Wenjie Hou}
%\email[E-mail: ]{wenjiehou@buaa.edu.cn} 
\affiliation{Institute of Fundamental and Frontier Sciences, University of Electronic Science and Technology of China, Chengdu 610054, China}
\affiliation{Taizhou Key Laboratory of Minimally Invasive Interventional Therapy and Artificial Intelligence, Taizhou, 317502, China}

\author{Hong Gu}
%\email[E-mail: ]{guh@szlab.ac.cn} 
\affiliation{Suzhou Laboratory, No.388, Ruoshui Street, SIP, Jiangsu, 215123, China}

\author{Xiaoying Wang}
%\email[E-mail: ]{wangxiaoying22@stu.xjtu.edu.cn} 
\affiliation{State Key Laboratory for Mechanical Behavior of Materials, School of Materials Science and Engineering,
Xi'an Jiaotong University, Xi'an 710049, China}

\author{Xuejie Li}
%\email[E-mail: ]{l940617225@stu.xjtu.edu.cn} 
\affiliation{State Key Laboratory for Mechanical Behavior of Materials, School of Materials Science and Engineering,
Xi'an Jiaotong University, Xi'an 710049, China}

\author{Jun Sun}
%\email[E-mail: ]{junsun@mail.xjtu.edu.cn} 
\affiliation{State Key Laboratory for Mechanical Behavior of Materials, School of Materials Science and Engineering,
Xi'an Jiaotong University, Xi'an 710049, China}
                
\author{Xiangdong Ding}
%\email[E-mail: ]{dingxd@mail.xjtu.edu.cn} 
\affiliation{State Key Laboratory for Mechanical Behavior of Materials, School of Materials Science and Engineering,
Xi'an Jiaotong University, Xi'an 710049, China}

%\author{Baowen Li}
%\email[E-mail: ]{libw@sustech.edu.cn}
%\affiliation{Department of Materials Science and Engineering, Department of Physics. Southern University of Science and Technology, Shenzhen, 518055, China}

\author{Zhibin Gao}
\email[E-mail: ]{zhibin.gao@xjtu.edu.cn}
\affiliation{State Key Laboratory for Mechanical Behavior of Materials, School of Materials Science and Engineering,
Xi'an Jiaotong University, Xi'an 710049, China}

\date{\today}
%---------------------------------------------------------------------
\begin{abstract}
Efficient and precise calculations of thermal transport properties and figure of merit, alongside a deep comprehension of thermal transport mechanisms, are essential for the practical utilization of advanced thermoelectric materials. In this study, we explore the microscopic processes governing thermal transport in the distinguished crystalline material Tl$_9$SbTe$_6$ by integrating a unified thermal transport theory with machine learning-assisted self-consistent phonon calculations. Leveraging machine learning potentials, we expedite the analysis of phonon energy shifts, higher-order scattering mechanisms, and thermal conductivity arising from various contributing factors like population and coherence channels. Our finding unveils an exceptionally low thermal conductivity of 0.31 W m$^{-1}$ K$^{-1}$ at room temperature, a result that closely correlates with experimental observations. Notably, we observe that the off-diagonal terms of heat flux operators play a significant role in shaping the overall lattice thermal conductivity of Tl$_9$SbTe$_6$, where the ultralow thermal conductivity resembles that of glass due to limited group velocities. Furthermore, we achieve a maximum $ZT$ value of 3.17 in the $c$-axis orientation for \textit{p}-type Tl$_9$SbTe$_6$ at 600 K, and an optimal $ZT$ value of 2.26 in the $a$-axis and $b$-axis direction for \textit{n}-type Tl$_9$SbTe$_6$ at 500 K. The crystalline Tl$_9$SbTe$_6$ not only showcases remarkable thermal insulation but also demonstrates impressive electrical properties owing to the dual-degeneracy phenomenon within its valence band. These results not only elucidate the underlying reasons for the exceptional thermoelectric performance of Tl$_9$SbTe$_6$ but also suggest potential avenues for further experimental exploration.

\end{abstract}

%---------------------------------------------------------------------

% PACS 2010 ALPHABETICAL INDEX
% https://publishing.aip.org/publishing/pacs/pacs-alphabetical-index
% PACS 2010 Regular Edition
% https://publishing.aip.org/publishing/pacs/pacs-2010-regular-edition
%\pacs{
%65.80.Ck,   %Thermal properties of graphene
%65.40.-b,    %Thermal properties of crystalline solids
%61.46.-w,    %Structure of nanoscale materials
%64.70.Nd,   %Structural transitions in nanoscale materials
%71.30.+h     %Metal-insulator transitions an                              d other electronic transitions
%64.70.K-    %Solid-solid transitions
%78.67.Pt     %Multilayers; superlattices; photonic structures; metamaterials
%64.70.-p    %Specific phase transitions
%64.70.K-    %Solid-solid transitions
%73.22.-f,    %Electronic structure of nanoscale materials and related
             %systems
%73.20.At    %Surface states, band structure, electron density of
%            %states
%81.05.Cy    %Elemental semiconductors
%81.05.Zx     %New materials: theory, design, and fabrication
%81.16.-c    %Methods of micro- and nanofabrication and processing
%81.30.-t    %Phase diagrams and microstructures developed by
%            %solidification and solid-solid phase transformations
% }

% Insert suggested keywords - APS authors don't need to do this

% \maketitle must follow title, authors, abstract, \pacs, and \keywords
\maketitle

%---------------------------------------------------------------------
\section{1. Introduction}
%Thermoelectric materials have the capability to convert heat into electrical energy without emitting pollutants~\cite{Snyder2008Complex,Zebarjadi2012Perspectives}. They have no movable components, generate no noise pollution, and are adjustable in size, making them indispensable in various fields such as waste heat recovery ~\cite{Baxter2009Nanoscale} and photovoltaic-photothermal cogeneration~\cite{Dallan2015Performance}. The conversion efficiency of thermoelectric materials can be determined by the figure of merit $ZT = S^2\sigma/(\kappa_e+\kappa_L)$, where $S$, $\sigma$, $\kappa_e$ and $\kappa_L$ are the Seebeck coefficient, electrical conductivity, electrical thermal conductivity and lattice thermal conductivity, respectively. To achieve good thermoelectric performance, a promising strategy involves enhancing electrical transport while simultaneously suppressing thermal transport~\cite{George2001Thermoelectrics}. However, it's challenging to achieve high electrical conductivity ($\sigma$) and low electronic thermal conductivity ($\kappa_e$) simultaneously. Therefore, ultralow lattice thermal conductivity ($\kappa_L$) becomes a crucial factor contributing to high thermoelectric efficiency. 

Thermoelectric materials have the capacity to transform heat into electrical energy without emitting 
pollutants~\cite{Snyder2008Complex,Zebarjadi2012Perspectives}. They lack movable components, generate no noise pollution, and are adjustable in size, rendering them indispensable in a multitude of fields, including waste heat recovery~\cite{Baxter2009Nanoscale} and 
photovoltaic-photothermal cogeneration~\cite{Dallan2015Performance}. The conversion efficiency of thermoelectric materials can be determined by the figure of merit $ZT = S^2\sigma/(\kappa_e+\kappa_L)$, where $S$, $\sigma$, $\kappa_e$ and $\kappa_L$ are the Seebeck coefficient, electrical conductivity, electrical thermal conductivity and lattice thermal conductivity, respectively. In order to achieve optimal thermoelectric performance, a promising strategy involves enhancing electrical transport while simultaneously suppressing thermal transport~\cite{George2001Thermoelectrics}. Nevertheless, it is challenging to achieve high electrical conductivity ($\sigma$) and low electronic thermal conductivity ($\kappa_e$) simultaneously. Consequently, ultralow lattice thermal conductivity ($\kappa_L$) becomes a crucial factor contributing to high thermoelectric efficiency.

Recent studies have identified several materials with relatively complex structures, including Cu$_{12}$Sb$_{4}$S$_{13}$~\cite{Xia2020Microscopic}, AgPbBiSe$_3$~\cite{Dutta2019Bonding}, Cs$_2$BiAgBr$_6$~\cite{Enamul2018Origin}, KCu$_5$Se$_3$~\cite{LiFan2023Overdamped}, and Cu$_7$PS$_6$~\cite{Shen2024Amorphous}. These materials exhibit potential for thermoelectric applications due to their ability to achieve ultralow thermal conductivity ($\kappa_L$). 
%Therefore, research on complex thermoelectric materials is not only fundamental interesting, but also is crucial for advancing thermoelectric technology. 
Therefore, research on complex thermoelectric materials is not only of fundamental interest but also of crucial importance for the advancement of thermoelectric technology.

The Tl-Te system contains numerous complex materials with ultralow $\kappa_L$.  For instance, the $\kappa$ of Ag$_8$Tl$_2$Te$_5$, Ag$_9$TlTe$_5$, Tl$_9$CuTe$_5$, Tl$_2$GeTe$_3$, TlInTe$_2$, etc. are all lower than 0.5 W m$^{-1}$ K$^{-1}$ at room temperature~\cite{Guo2014Thermoelectric,Yamanaka2006MATERIALS}. Moreover, the $\kappa$ of the promising thermoelectric materials Tl$_9$BiTe$_6$ and Tl$_9$SbTe$_6$ was reported to be 0.5 W m$^{-1}$ K$^{-1}$ and 0.7 W m$^{-1}$ K$^{-1}$, respectively, at room temperature~\cite{Guo2014Thermoelectric}. Although experiments have shown that Tl$_9$BiTe$_6$ exhibits lower thermal conductivity, it also exhibits high resistivity, which significantly reduces the power factor and constrains its further enhancement as an additional component. %Because even with little Tl$_9$BiTe$_6$ addition, it can greatly reduce thermoelectric performance of the material~\cite{Teubner2001Optimization}. 
Even with minimal Tl$_9$BiTe$_6$ incorporation, it can markedly diminish the thermoelectric performance of the material~\cite{Teubner2001Optimization}.

Compound Tl$_9$SbTe$_6$ represents the most closely related material to Tl$_9$BiTe$_6$ within the Tl$_5$Te$_3$ family.
%is the closest material to Tl$_9$BiTe$_6$ in the Tl$_5$Te$_3$ family. 
Its thermal conductivity is comparable to that of Tl$_9$BiTe$_6$, yet its electrical conductivity is markedly superior, as evidenced by the 
experimental results~\cite{Guo2014Thermoelectric}. 
Although there have been experimental studies on thermoelectric properties~\cite{Guo2013Enhanced,Guo2014Thermoelectric,Guo2015The} and analysis of electronic structure~\cite{Tao2011Physical}, there is still a lack of in-depth research on its low $\kappa$ and high thermoelectric efficiency due to its complex structure. Further comprehensive research will help us better understand and design excellent thermoelectric materials based on the complex Tl-Te system.

Despite the significant advancements in measuring ultralow $\kappa$ in experiments, theoretical prediction and explanation of ultralow $\kappa$ in complex compounds still remain challenging. Recent studies have explored anharmonic effects as potential sources of discrepancies of $\kappa$ between the experiment and theory observed in highly anharmonic crystals~\cite{Xia2020Particlelike,wang2023role,Tadano2018Quartic,Xia2020Microscopic,Zheng2022Anharmonicity,Yue2024Ultralow,Zheng2024Unravelling}. 
These studies have revealed that lattice anharmonicity significantly influences the phonon linewidth and lattice thermal conductivity~\cite{Tadano2015consistent}. While the conventional Peierls-Boltzmann framework is adequate for describing phonon transport in systems with well-defined phonon modes, it encounters challenges in highly anharmonic systems~\cite{Tadano2015consistent,Zhao2021Lattice}. Therefore, it becomes imperative to consider the temperature-dependent frequencies evaluated by the self-consistent phonon (SCPH) approach in elucidating thermal transport in complex compounds~\cite{Xia2020Throughput,Kang2019Intrinsic}.

Moreover, a discrepancy exists between the experimentally measured and theoretically predicted $\kappa$ by the conventional Peierls-Boltzmann transport equation~\cite{Simoncelli2019}. Consequently, it becomes necessary to consider the influence of wavelike interbranch tunneling of coherence, originating from off-diagonal terms of the heat flux operator~\cite{Simoncelli2019}, to comprehensively explain this phenomenon. Overall, to accurately predict thermal transport properties in complex thermoelectric materials, the higher-order force constants, i.e., fourth-order force constants, and coherent phonons are required to explicitly evaluate the phonon energy shifts and scattering rates.
Overall, to accurately predict thermal transport properties in complex thermoelectric materials, it is necessary to explicitly evaluate the phonon energy shifts and scattering rates using the higher-order force constants i.e., fourth-order force constants, and coherent phonons.

Despite the significant achievements in DFT-based thermal conductivity calculations~\cite{WuLI2014ShengBTE}, the computational expense of calculating thermal conductivity considering higher-order phonon scatterings is so high that it is often prohibitive~\cite{Han2022FourPhonon,Tong2023glass}. 
The use of machine Learning Potential (MLP) can significantly reduce the cost of calculating force constants while maintaining considerable computational accuracy~\cite{Bohayra2021Accelerating}. Machine learning potential (MLP) is a potential that employs machine learning algorithms to describe the atomic potential energy surface and predict the energy and force based on the atomic environment. Unlike machine learning models (ML-Models)~\cite{Wan2019Materials}, MLP can be combined with BTE methods to provide detailed phonon information, thereby enabling a more comprehensive understanding of the mechanism of thermal transport in the large system. 
%And it can calculate complex systems under various conditions, such as defects, disorder, anharmonicity, and other complex systems that are difficult to calculate using DFT methods. 
Furthermore, it is capable of calculating complex systems under a multitude of conditions, including those involving defects, disorder, anharmonicity, and other complex systems that are challenging to calculate using DFT methods.

Shapeev et al. developed another machine learning potential function, the moment tensor potential (MTP)~\cite{Novikov2021MLIP,Shapeev2016Moment}, which achieves a good balance between computational accuracy and computational complexity. 
MTP can obtain high-precision second-order and third-order force constants. According to the previous research, MTP can also obtain reliable fourth-order force constants, which can accurately calculate the $\kappa_L$ containing four-phonon scattering~\cite{Ouyang2022Accurate}. Consequently, based on the fourth-order anharmonicity~\cite{Han2022FourPhonon}, we have modified and developed corresponding interface using the MTP to obtain fourth-order force constants in an efficient manner.

%In this work, we comprehensively investigated the thermal transport and thermoelectric properties of benchmark compund Tl$_9$SbTe$_6$ by integrating machine-learning based anharmonic lattice dyanmics with an unified theory of thermal transport and momentum relaxation time approximation, including: (1) Constructing machine-learning potential with DFT-level accuracy for complex crystalline Tl$_9$SbTe$_6$, (2) anharmonically renormalizing phonons using SCPH approach, (3) evaluating thermal transport considering both particle-like and wave-like phonon transport channels, (4) employing AMSET to systematically study the material's electrical transport properties, (5) exploring the trend of the thermoelectric property $ZT$ value with carrier concentration and temperature variation.  

In this study, we conducted a comprehensive investigation of the  thermal transport and thermoelectric properties of the benchmark compound Tl$_9$SbTe$_6$. To this end, we integrated machine-learning based anharmonic lattice dyanmics with an unified theory of thermal transport and momentum relaxation time approximation. This integration included: (1) Construction of a machine-learning potential with DFT-level accuracy for the complex crystalline Tl$_9$SbTe$_6$, (2) anharmonically renormalizing phonons using the SCPH approach, (3) evaluation of thermal transport considering both particle-like and wave-like phonon transport channels, (4) application of distinct scattering mechanisms, including acoustic deformation potential (ADP), ionized impurity (IMP) scattering, and polar
optical phonon (POP) scattering, to systematically study the electrical transport properties, (5) exploration of the trend of the thermoelectric property $ZT$ value with carrier concentration and temperature variation.

%\section{2. Computational Methods}

%results
\section{2. Results and discussion}
\subsection{2.1 Crystal structure and machine learning potential}
%\textbf{2.1 Crystal structure and machine learning potential}

%===========< FIGURE 1 >=========================================
\begin{figure*}%[htp]
\includegraphics[width=2.0\columnwidth]{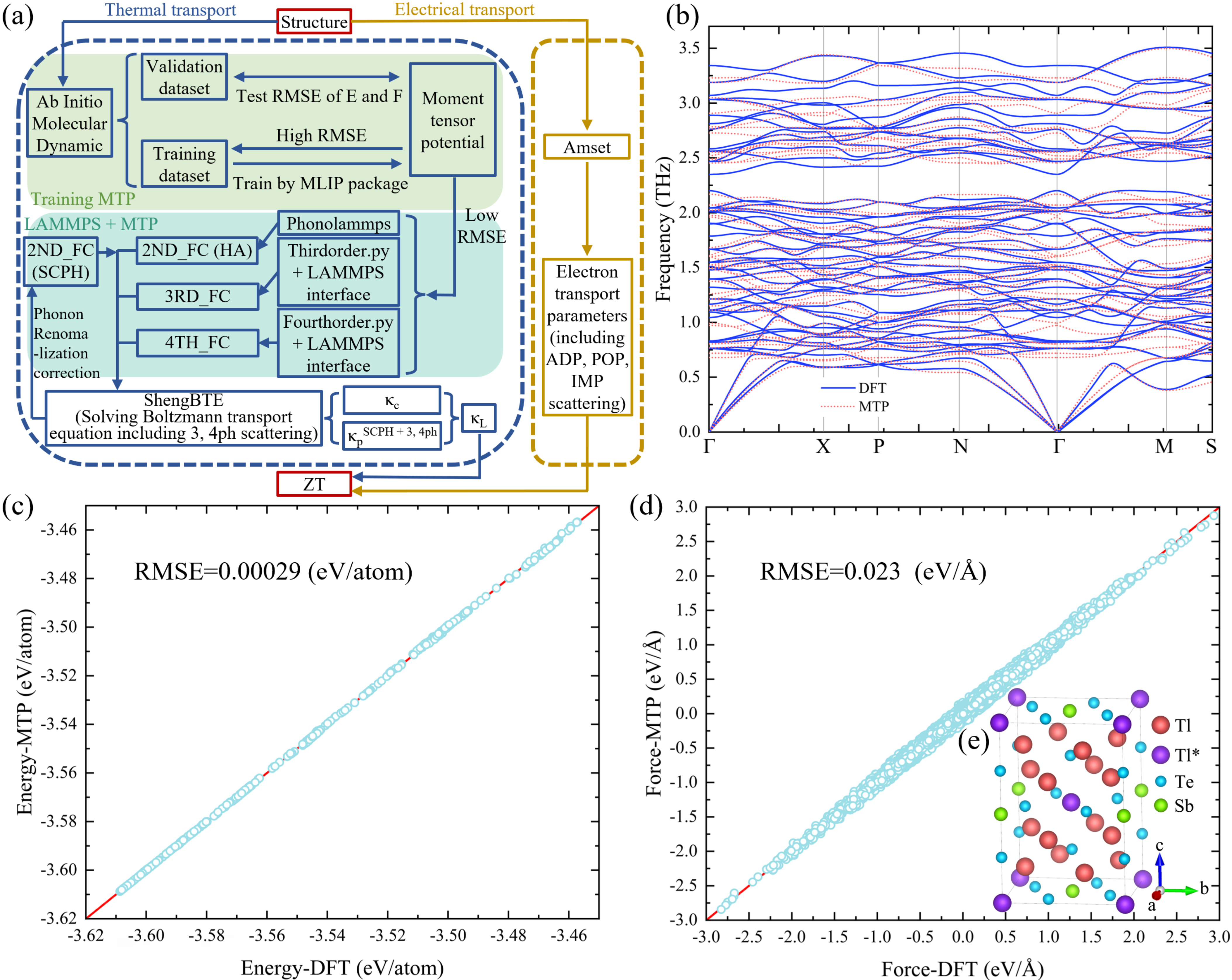}
\caption{(a) The workflow of the thermoelectric property calculations. (b) Phonon dispersion of Tl$_9$SbTe$_6$ calculated by Moment tensor potential (MTP) and DFT. A comparisons of (c) the energy per atom (d) and the atomic force between MTP and DFT calculations for Tl$_9$SbTe$_6$. (e) The conventional cell structure of Tl$_9$SbTe$_6$. The red and purple atoms represent Thallium (Tl) atoms in different chemical environments. The atoms represented by the blue color are Tellurium (Te) atoms, while the green ones are Antimony (Sb) atoms.}
\label{fig1}
\end{figure*}
%===========< FIGURE 1 >=========================================

%===========< FIGURE 2 >=========================================
\begin{figure*}%[htp]
\includegraphics[width=2.0\columnwidth]{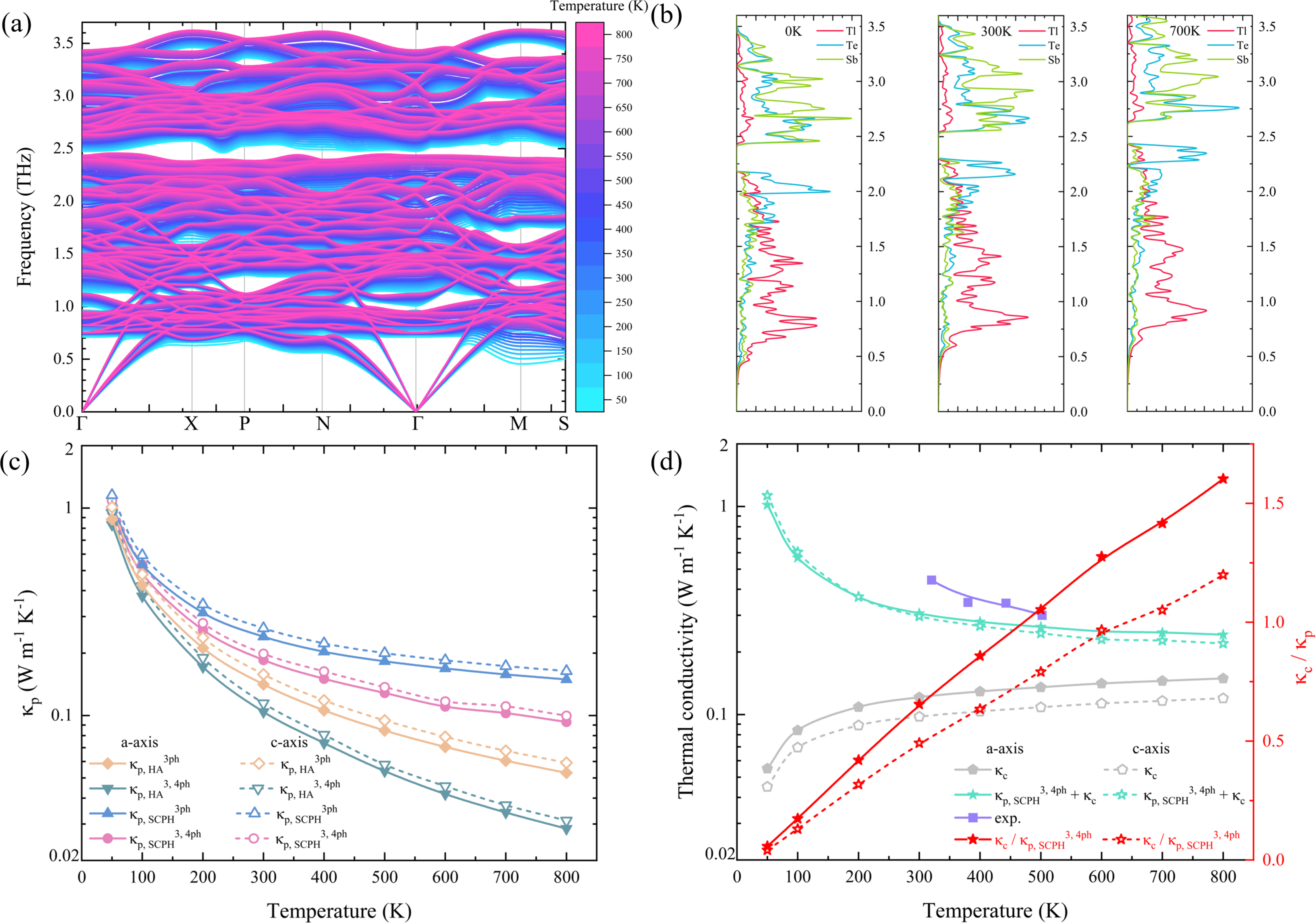}
%\vspace{-2mm}
\caption{(a) The temperature-dependent phonon spectrum from T = 50 K to 800 K for Tl$_9$SbTe$_6$. (b) Projected phonon density of states (PDOS) of Tl$_9$SbTe$_6$ at 0 K, 300 K, and 700 K.  (c) Phonon populations’ thermal conductivity ($\kappa_p$) as a function of the temperature of Tl$_9$SbTe$_6$. (d) Intrinsic lattice thermal conductivity ($\kappa_L$) as a function of the temperature of Tl$_9$SbTe$_6$. 
The ratio $\kappa_c/\kappa_p$ represents $\kappa_{p, SCPH}^{3, 4ph}$. The solid lines represent the $a$-axis direction, and the dashed lines represent the $c$-axis direction. %The rectangle symbol represents the $\kappa_L$ considering three-phonon (3ph) scattering processes, while the circle symbol represents the $\kappa_L$ considering both the 3ph and the four-phonon (4ph) scattering processes. 
%There are three acoustic branches and fourty-five optical branches.
\label{fig2}}
\end{figure*}
%===========< FIGURE 2 >=========================================

%===========< FIGURE 3 >=========================================
\begin{figure*}%[htp]
\includegraphics[width=2.0\columnwidth]{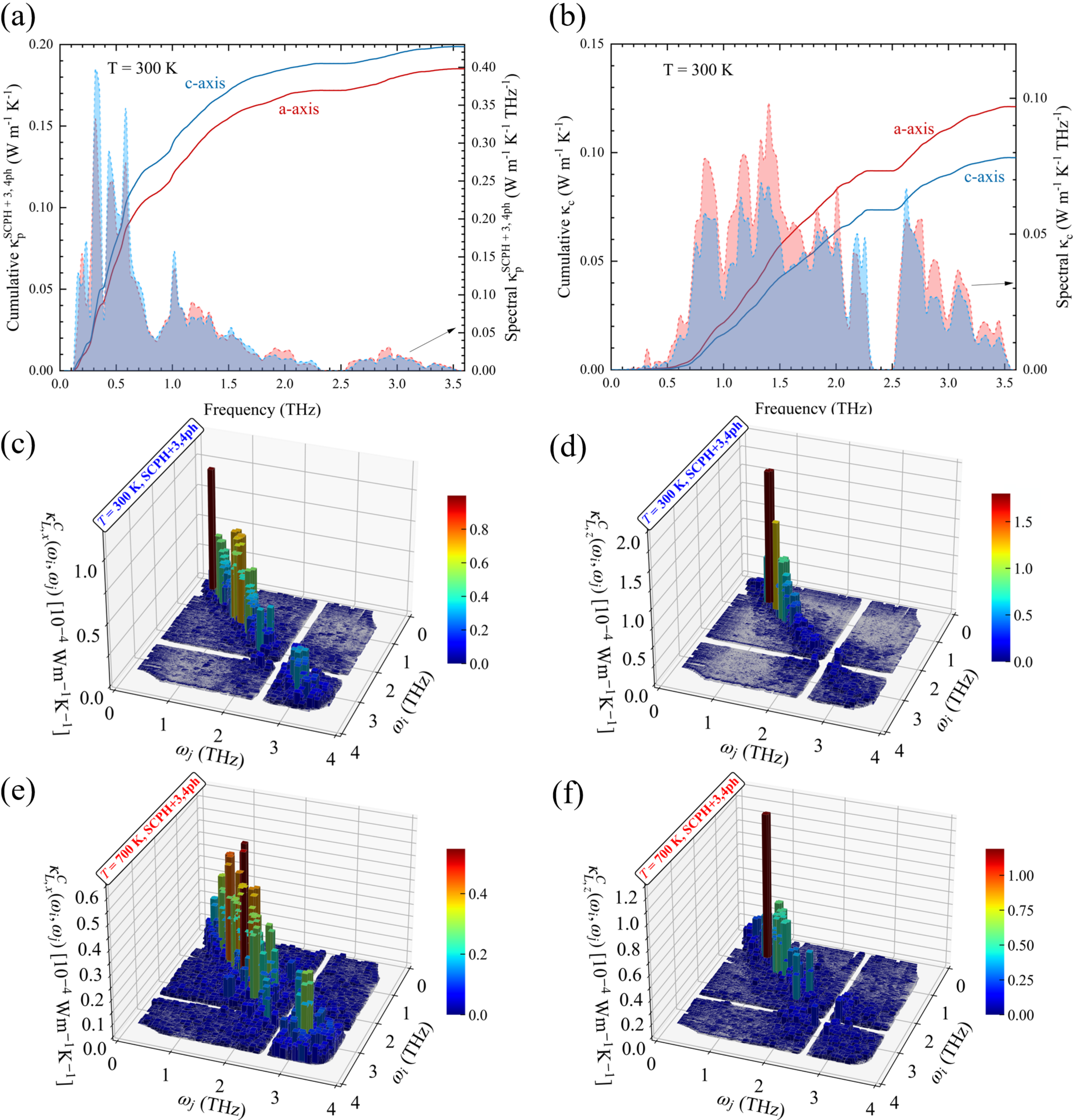}
\caption{(a) Calculated spectral and cumulative $\kappa_p$ using the SCPH+3,4ph model at 300 K. (b) Calculated spectral and cumulative coherence thermal conductivity at 300 K. (c) Three-dimensional visualization model $\kappa_c$($\omega_i$, $\omega_j$) of the contributions to the coherence thermal conductivity in the $a$-axis direction at 300 K. 
(d) The same as (c) but in $c$-axis direction. 
(e) The same as (c) but at 700 K. 
(f) The same as (d) but at 700 K.
\label{fig3}}
\end{figure*}
%===========< FIGURE 3 >=========================================

%The lattice constants of the optimized structure of Tl$_9$SbTe$_6$ by using LDA potentials are $a$ = $b$ = 8.63 \AA, and $c$ = 12.77 \AA. They are in consistent with the experimental values which is $a$ = $b$ = 8.830 \AA~ and $c$ = 13.013 \AA ~\cite{Guo2014Thermoelectric,Tao2011Physical}. As shown in Fig.~\ref{fig1}(b), the phonon spectrum of the LDA functionals is free of imaginary frequencies. The primitive cell contains 16 atoms: one Sb, six Te, and nine Tl. The Tl atoms are divided into eight Tl and one Tl*, based on their different chemical environments and locations, as shown in Fig.~\ref{fig1}(e). 

The optimized lattice constants of Tl$_9$SbTe$_6$, calculated using LDA potentials, are $a$ = $b$ = 8.63 \AA, and $c$ = 12.77 \AA. These values are consistent with the experimental values, which are $a$ = $b$ = 8.830 \AA~ and $c$ = 13.013 \AA~\cite{Guo2014Thermoelectric,Tao2011Physical}. As illustrated in Fig.~\ref{fig1}(b), the phonon spectrum of the LDA functional is devoid of imaginary frequencies, indicating the dynamical stability of Tl$_9$SbTe$_6$. The primitive cell comprises 16 atoms: one Sb, six Te, and nine Tl. Tl atoms are divided into eight Tl and one Tl*, based on their deisparate chemical environments and locations, as depicted in Fig.~\ref{fig1}(e). 

%Fig.~\ref{fig1}(a) depicts the workflow of our work. In this work, we obtain the training dataset from Ab Initio Molecular Dynamic 
%(AIMD) simulations using the isobaric-isothermal (NPT) ensemble. We train the MTP using the MLIP package~\cite{Novikov2021MLIP}, and the %training principles are detailed in the Supplementary Informations (SI) [\textcolor{red}{reference}]. After obtaining the MTP, we test it on 
%training principles are detailed in the Supplementary Information. After obtaining the MTP, we test it on 
%the validation dataset. We use the MTP to determine the energy and atomic forces of each structure in the validation dataset and calculate the root-mean-square error (RMSE), as shown in Figs.~\ref{fig1}(c-d). The RMSE of energy per atom and atomic force is 0.00029 eV/atom and 0.023 eV/Å, respectively, indicating that the accuracy of the trained MTP is sufficient for calculating thermodynamic properties~\cite{Ouyang2022Accurate}. Moreover, as shown in Fig.~\ref{fig1}(b), the phonon spectrum calculated using the moment tensor potential (MTP) is in good agreement with the DFT results. %The phonon spectrum does not exhibit any imaginary frequencies, indicating that Tl$_9$SbTe$_6$ is dynamically stable.
%Note that since the phonon spectrum of the PBE functionals shows an imaginary frequency as demonstrated in the Supplementary Material, we adopt the LDA functionals for the subsequent calculations.

Fig.~\ref{fig1}(a) depicts the workflow of our work. In this work, we obtain the training dataset from \textit{Ab Initio} Molecular Dynamics (AIMD) using the isobaric-isothermal (NPT) ensemble. We then train the MTP and the training principles are detailed in the Supplementary Information~\cite{Novikov2021MLIP}. After obtaining the MTP, we test it on the validation dataset. 

The MTP is employed to ascertain the energy and atomic forces of each configuration in the validation dataset, with the resulting root-mean-square error (RMSE) being presented in Figs.~\ref{fig1}(c-d). The RMSE of energy per atom and atomic force is 0.00029 eV/atom and 0.023 eV/\AA, respectively. These values indicate that the trained MTP is sufficiently accurate for calculating thermodynamic properties~\cite{Novikov2021MLIP,Ouyang2022Accurate}. Furthermore, as shown in Fig.~\ref{fig1}(b), the phonon spectrum calculated by MTP is 
%using the MTP method 
in good agreement with the DFT results.

\subsection{2.2 Anharmonic lattice dynamics and thermal transport properties}
%\textbf{2.2 Anharmonic lattice dynamics and thermal transport properties}

%Using the interatomic force constants (IFCs) obtained from MLP, we next examine the impact of fourth-order anharmonicity on the phonon spectra at finite temperatures, as depicted in Figs.~\ref{fig2}(a-b). Clearly, an overall phonon hardening phenomenon is observed in crystalline Tl$_9$SbTe$_6$ as the temperature increases from 50 to 800 K. This observation suggests strong lattice anharmonicity in Tl$_9$SbTe$_6$ and highlights the importance of properly treating lattice anharmonicity in lattice dynamics~\cite{Zheng2022Anharmonicity,Tadano2015consistent,wang2023role}. More specifically, the low-frequency phonons with frequency less than 1.5 THz are dominated by Tl atoms, as revealed by the projected phonon density of states (PDOS) in the Fig.~\ref{fig2}(b). The strong phonon hardening observed in the Tl-dominated modes can be attributed to the large atomic displacements resulting from their loose bonding [see Fig. S1 in SI]. Additionally, we observe that Tl atoms contribute to the low-lying flattened phonon bands in the frequency range of 0.7 to 1.0 THz. This suggests strong phonon scattering, which suppresses the primary heat carriers, i.e., acoustic modes~\cite{Li2015Ultralow,Zheng2022Effects,wang2024anomalous}. On the other hand, the optical modes with higher energies exhibit less increase in stiffness corresponding to the majority of Te and Sb vibrations.

The impact of forth-order anharmonicity on the phonon spectra at finite temperatures is next examined, as depicted in Figs.~\ref{fig2}(a-b), using the interatomic force constants (IFCs) obtained from MLP. It is evident that overall phonon hardening phenomenon is observed in crystalline Tl$_9$SbTe$_6$ as the temperature increases from 50 to 800 K. This observation suggests the presence of strong lattice anharmonicity in Tl$_9$SbTe$_6$ and highlights the importance of properly treating lattice anharmonicity in lattice dynamics~\cite{Zheng2022Anharmonicity,Tadano2015consistent,wang2023role,feng2024relation}. In particular, the low-frequency phonons with a frequency below 1.5 Thz are predominantly influenced by the Tl atoms, as evidenced by the projected phonon density of states (PDOS) in Fig.~\ref{fig2}(b). The pronounced phonon hardening observed in the Tl-dominated modes can be attributed to the significant atomic displacements resulting from their loose bonding, as illustrated in Fig. S1 in the supplementary information. Furthermore, it can be observed that Tl atoms contribute to the low-lying flattened phonon bands in the frequency range of 0.7 to 1.0 THz. This suggests that strong phonon scattering is occurring, which suppresses the primary heat carriers, i.e., acoustic phonon modes~\cite{Li2015Ultralow,Zheng2022Effects,wang2024anomalous}. In contrast, the optical modes with higher energies exhibit less increase in stiffness corresponding to the majority of Te and Sb vibrations. 

%With the zero-K/finite-temperature IFCs at hand, we proceed to calculate the temperature-dependent $\kappa_{p}$ using three different levels of theory in Figs.~\ref{fig2}(c-d). The thermal conductivity ($\kappa_L$) is anisotropic; in the $a$-axis or $b$-axis direction, the $\kappa_p$ is approximately 10 percent smaller than that in the $c$-axis direction. Using the lowest level of thermal transport theory, namely, the HA+3ph model, $\kappa_{p, HA}^{3ph}$ in the $a$-axis is 0.88, 0.14, and 0.053 W m$^{-1}$ K$^{-1}$ at 50, 300, and 800 K, respectively. As mentioned earlier, anharmonic phonon renormalization is non-negligible in predicting finite-temperature $\kappa_L$ in complex compounds. With the additional effect of phonon energy shifts, we advance the HA+3ph model to a more accurate SCPH+3ph model, which gives $\kappa_{p, SCPH}^{3ph}$ values of 1.02, 0.24, and 0.15 W m$^{-1}$ K$^{-1}$ at 50, 300, and 800 K, respectively, in the $a$-axis direction. These values are 16\%, 71\%, and 183\% larger than $\kappa_{p, HA}^{3ph}$, indicating significant anharmonic phonon renormalization in Tl$_9$SbTe$_6$, with a substantial increase as temperature rises. This trend is also confirmed by Fig.~\ref{fig2}(a), which shows significant phonon frequency shifts with increasing temperature. 

With the zero-K/finite-temperature IFCs at hand, we proceed to calculate the temperature-dependent $\kappa_{p}$ using three different levels of theory, as shown in Figs.~\ref{fig2}(c-d). It is observed that the lattice thermal conductivity ($\kappa_L$) is anisotropic, with the $\kappa_p$ approximately 10\% smaller in the $a$-axis or $b$-axis direction than in the $c$-axis direction. The lowest level of thermal transport theory, namely the HA+3ph model, yields the following values for $\kappa_{p, HA}^{3ph}$ in the $a$-axis at 50, 300, and 800 K: 0.88, 0.14, and 0.053 W m$^{-1}$ K$^{-1}$, respectively. As previously stated, anharmonic phonon renormalization is a significant factor in the prediction of finite-temperature $\kappa_L$ in complex compounds. With the additional effect of phonon energy shifts, we advance the HA+3ph model to a more accurate SCPH+3ph model, which gives $\kappa_{p, SCPH}^{3ph}$ values of 1.02, 0.24, and 0.15 W m$^{-1}$ K$^{-1}$ at 50, 300, and 800 K, respectively, in the $a$-axis direction. These values are 16\%, 71\%, and 183\% larger than $\kappa_{p, HA}^{3ph}$, indicating significant anharmonic phonon renormalization in Tl$_9$SbTe$_6$, with a substantial increase as temperature rises. This trend is also confirmed by Fig.~\ref{fig2}(a), which shows significant phonon frequency shifts with increasing temperature. 

%Meanwhile, the quartic anharmonicity not only induces large phonon frequency shifts but also results in strong 4ph scatterings. Therefore, we further include additional 4ph scatterings to obtain $\kappa_{p, SCPH}^{3, 4ph}$ values of 0.96, 0.19, and 0.093 W m$^{-1}$ K$^{-1}$ at 50, 300, and 800 K, respectively, in the $a$-axis direction [see Fig.~\ref{fig2}(c)]. These values represent decreases of 6\%, 21\%, and 38\% compared with $\kappa_{p, HA}^{3ph}$, indicating the importance of the 4ph scattering processes~\cite{wang2023role}. Interstingly, the difference between the $\kappa_{p, HA}^{3ph}$ and $\kappa_{p, HA}^{3, 4ph}$ is 6\%, 26\%, and 46\% larger than difference between the anharmonic phonon renormalized ones, which means that phonon renormalization and 4ph scattering induced by quartic anharmonicity have the opposite effect on thermal transport. 

In addition, the quartic anharmonicity not only induces significant shifts in phonon frequencies but also results in pronounced 4ph scatterings. Consequently, additional 4ph scatterings were included to obtain $\kappa_{p, SCPH}^{3, 4ph}$ values of 0.96, 0.19, and 0.093 W m$^{-1}$ K$^{-1}$ at 50, 300, and 800 K, respectively, in the $a$-axis direction (see Fig.~\ref{fig2}(c)).  These values represent decreases of 6\%, 21\%, and 38\% compared with $\kappa_{p, HA}^{3ph}$, indicating the importance of the 4ph scattering processes~\cite{wang2023role,wang2024anomalous,wang2024revisiting}. It is noteworthy that the discrepancy between the $\kappa_{p, HA}^{3ph}$ and $\kappa_{p, HA}^{3, 4ph}$ is 6\%, 26\%, and 46\% greater than that between the anharmonic phonon renormalized ones. This implies that phonon renormalization and 4ph scattering induced by quartic anharmonicity exert an inverse influence on thermal transport. 

%However, $\kappa_{p, SCPH}^{3, 4ph}$ is still much lower than the experimentally measured $\kappa_L$, as shown in Fig.~\ref{fig2}(d), suggesting that coherence contributions are non-negligible. By further considering the off-diagonal terms of heat flux operators, the total $\kappa_L$ is significantly enhanced by incorporating both $\kappa_{p, SCPH}^{3, 4ph}$ and $\kappa_c$~\cite{Simoncelli2019}. Here $\kappa_c$ can be fomulated as,

Nevertheless, $\kappa_{p, SCPH}^{3, 4ph}$ remains considerably lower than the experimentally measured $\kappa_L$, as illustrated in Fig.~\ref{fig2}(d), indicating that coherence contributions are not negligible. By further considering the off-diagonal terms of heat flux operators, the total $\kappa_L$ significantly enhanced by incorporating both $\kappa_{p, SCPH}^{3, 4ph}$ and $\kappa_c$~\cite{Simoncelli2019}. In this context, $\kappa_c$ can be expressed as, 
\begin{equation}
\begin{aligned}
\label{equation1} 
\kappa_{c}= & \frac{\hbar^{2}}{k_{B} T^{2} V N_{0}} \sum_{\boldsymbol{q}} \sum_{s \neq s^{\prime}} \frac{\omega_{\boldsymbol{q}}^{s}+\omega_{\boldsymbol{q}}^{s^{\prime}}}{2} \boldsymbol{v}_{\boldsymbol{q}}^{s, \boldsymbol{s}^{\prime}} \boldsymbol{v}_{\boldsymbol{q}}^{\boldsymbol{s}^{\prime}, s} \\
& \times \frac{\omega_{\boldsymbol{q}}^{s} n_{\boldsymbol{q}}^{s}\left(n_{\boldsymbol{q}}^{s}+1\right)+\omega_{\boldsymbol{q}}^{s^{\prime}} n_{\boldsymbol{q}}^{s^{\prime}}\left(n_{\boldsymbol{q}}^{s^{\prime}}+1\right)}{4\left(\omega_{\boldsymbol{q}}^{s^{\prime}}-\omega_{\boldsymbol{q}}^{s}\right)^{2}+\left(\Gamma_{\boldsymbol{q}}^{s}+\Gamma_{\boldsymbol{q}}^{s^{\prime}}\right)^{2}} \\
& \times\left(\Gamma_{\boldsymbol{q}}^{s}+\Gamma_{\boldsymbol{q}}^{s^{\prime}}\right),
\end{aligned}
\end{equation}
where $\hbar$ is the reduced Planck constant, $k_{B}$ is Boltzmann constant, $V$ is the primitive cell volume, and $N_{0}$ is the total number of sampled phonon vectors. 
%
%In Fig.~\ref{fig2}(d), the ratio $\kappa_c/\kappa_p$ is higher than 0.5 at 300 K and increases significantly with temperature. Specifically, in the $a$-axis direction, the ratio of $\kappa_c/\kappa_p$ grows to 1.60 at 800 K, which is 2.44 times larger than the 0.65 observed at 300 K. This observation suggests that $\kappa_c$ is the dominant contribution to $\kappa_L$ in crystalline Tl$_9$SbTe$_6$ at temperatures higher than 300 K. The dominant role of coherence contribution in thermal transport was also observed in other complex compounds such as Cs$_2$AgBiBr$_6$~\cite{Zheng2024Unravelling}, Cu$_{12}$Sb$_{4}$S$_{13}$~\cite{Xia2020Microscopic} and CsPbBr$_3$ ~\cite{He2018High}. Although the heat transport phenomena in Tl$_9$SbTe$_6$ exhibit similarities to those in glasses, the strong anharmonicity is the key factor contributing to coherence conductivity, whereas in glasses, the off-diagonal contributions arise from structural disorder. 

As illustrated in Fig.~\ref{fig2}(d), the ratio $\kappa_c/\kappa_p$ is greater than 0.5 at 300 K and exhibits a pronounced increase with temperature. Specifically, in the $a$-axis direction, he ratio of $\kappa_c/\kappa_p$ reaches 1.60 at 800 K, which is 2.44 times larger than the 0.65 observed at 300 K. This observation suggests that $\kappa_c$ is the dominant contribution to $\kappa_L$ in crystalline Tl$_9$SbTe$_6$ at temperatures above 600 K. The dominant role of coherence contribution in thermal transport was also observed in other complex compounds such as Cs$_2$AgBiBr$_6$~\cite{Zheng2024Unravelling}, Cu$_{12}$Sb$_{4}$S$_{13}$~\cite{Xia2020Microscopic}, CsPbBr$_3$ ~\cite{He2018High}, and even simple cubic CsCl~\cite{wang2024revisiting}. Although the heat transport phenomena in Tl$_9$SbTe$_6$ exhibit similarities to those in glasses, the strong phonon anharmonicity is the key factor contributing to coherent thermal conductivity. In contrast, in glasses, the off-diagonal contributions arise from structural disorder. 

%After incorporating $\kappa_c$, the total $\kappa_L$ is approximately 0.27 W m$^{-1}$ K$^{-1}$ and 0.26 W m$^{-1}$ K$^{-1}$ at 400 K and 500 K, respectively. This agrees with experimental measurements, where $\kappa_L$ is around 0.34 and 0.30 W m$^{-1}$ K$^{-1}$ at 400 K and 500 K, respectively~\cite{Guo2014Thermoelectric}. The experimentally measured $\kappa_L$ is calculated as the difference of two similar experimentally determined value which are total thermal conductivity ($\kappa$) and electronic thermal conductivity ($\kappa_e$), meaning that the relative error is non-negligible~\cite{Guo2014Thermoelectric}, explaining the little difference between our $\kappa_L$ and the experimental predicted ones. Also, we found that the dependence of $\kappa$ to the temperature approximately follow T$^{-1/3}$, which indicating the glass-like behaviour of the Tl$_9$SbTe$_6$.

%===========< FIGURE 4 >=========================================
\begin{figure*}%[htp]
\includegraphics[width=2.0\columnwidth]{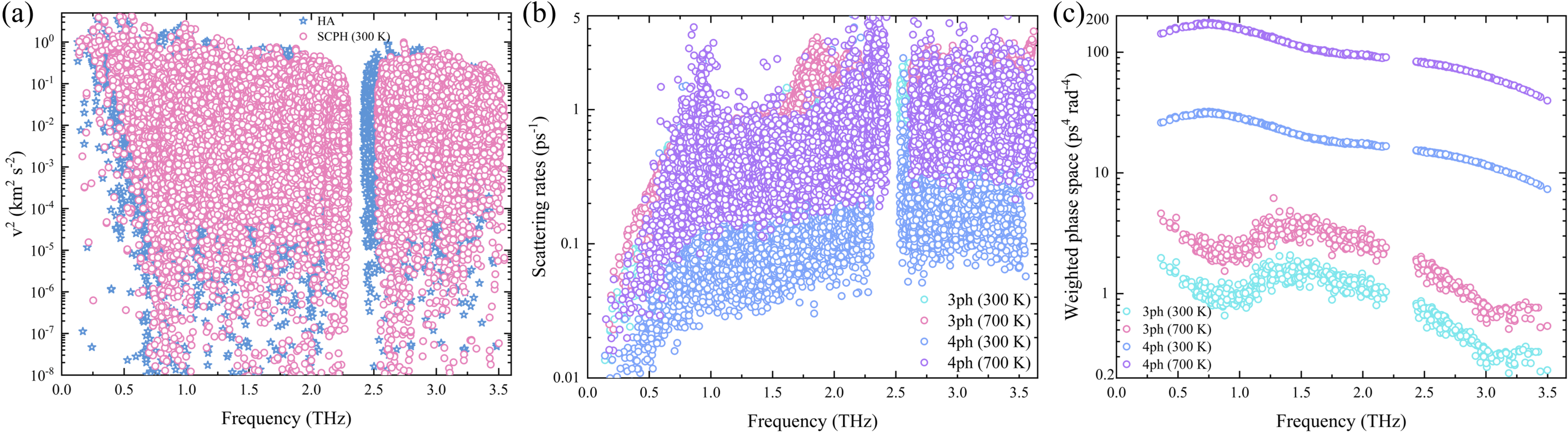}
%\vspace{-2mm}
\caption{(a) Squared phonon velocities of Tl$_9$SbTe$_6$ in the harmonic approximation and anharmonic renormalization approximation at 300 K.
%Clearly, the anharmonic phonon renormalization shows negligible effect on phonon group velocity. 
(b) Phonon scattering rates of 3ph and 4ph for Tl$_9$SbTe$_6$ at 300 K and 700 K. 
(c) Weighted phonon scattering phase space of 3ph and 4ph for Tl$_9$SbTe$_6$ at 300 K and 700 K.
%Squared phonon velocities of Tl$_9$SbTe$_6$ along the (a) $a$-axis and (b) $c$-axis with mode resolution. (c) Weighted phonon scattering phase space of 3ph and 4ph for Tl$_9$SbTe$_6$ at 300 K and 700 K. (d) Grüneisen parameter ($\gamma$) as a function of the phonon frequency. ZA (red), TA (purple), and LA (light blue) indicate the out-of-plane, in-plane transverse, and in-plane longitudinal phonon modes, respectively. (e) Phonon scattering rates of 3ph and 4ph for Tl$_9$SbTe$_6$ at 300 K and 700 K.\textcolor{red}{exchange the order of Fig. (b) and (c) and change it in the main content correspondingly.}
\label{fig4}}
\end{figure*}
%===========< FIGURE 4 >=========================================

%===========< FIGURE 5 >=========================================
\begin{figure*}%[htp]
\includegraphics[width=1.75\columnwidth]{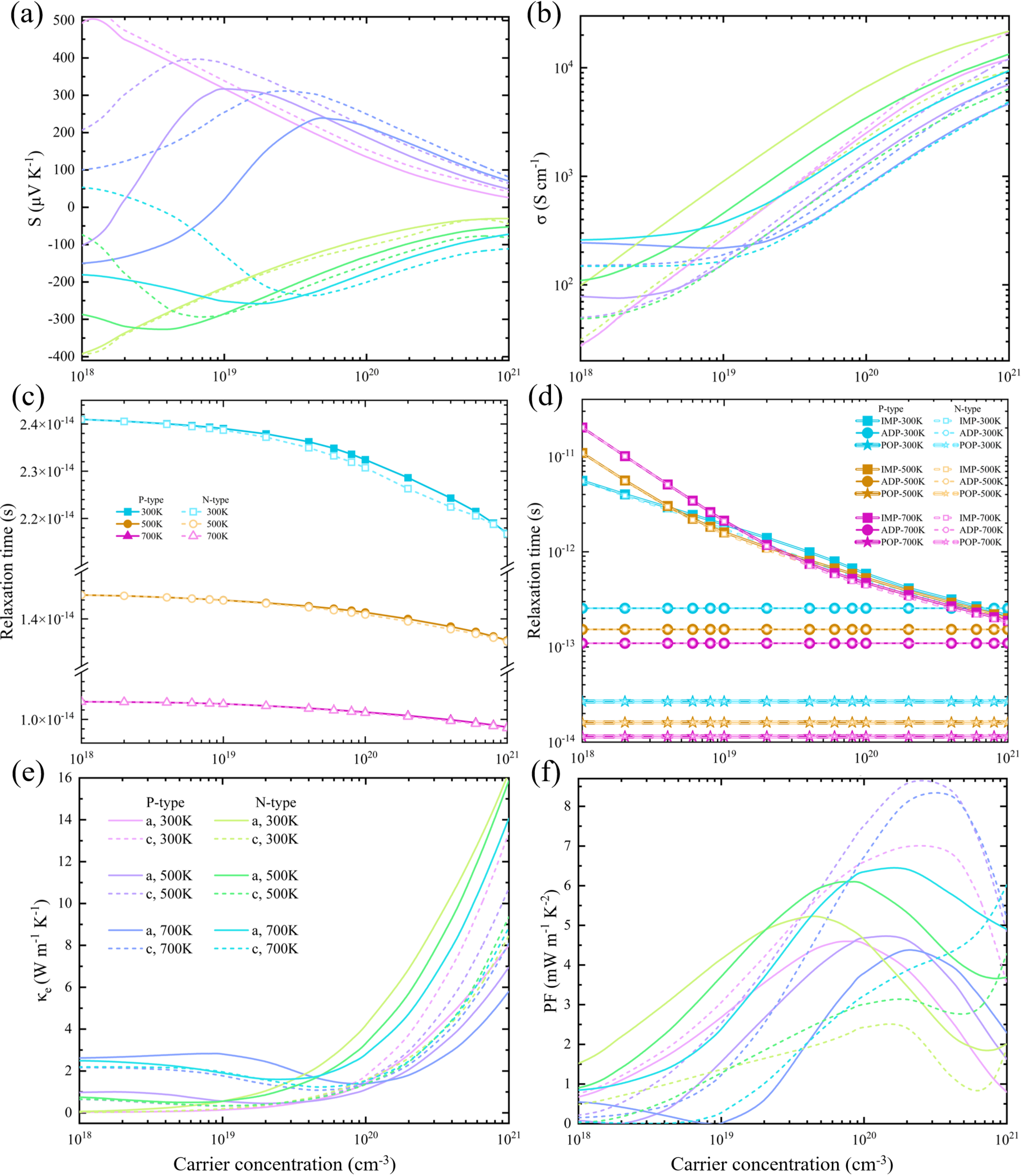}
%\vspace{-2mm}
\caption{The temperature-dependent (a) Seebeck coefficient \textit{S}, (b) electrical conductivity $\sigma$, (d) electronic thermal conductivity $\kappa_e$, and (e) the power factor ($PF=S^2\sigma$) of the Tl$_9$SbTe$_6$ were calculated as a function of carrier concentration (holes and electrons) along different axes at 300 K, 500 K and 700 K, respectively. 
(c) and (d) represent the electronic relaxation time for distinct scattering mechanisms, including acoustic deformation potential (ADP), ionized impurity (IMP) scattering, and polar optical phonon (POP) scattering, respectively.
\label{fig5}}
\end{figure*}
%===========< FIGURE 5 >========================================

%===========< FIGURE 6 >=========================================
\begin{figure*}%[htp]
\includegraphics[width=1.75\columnwidth]{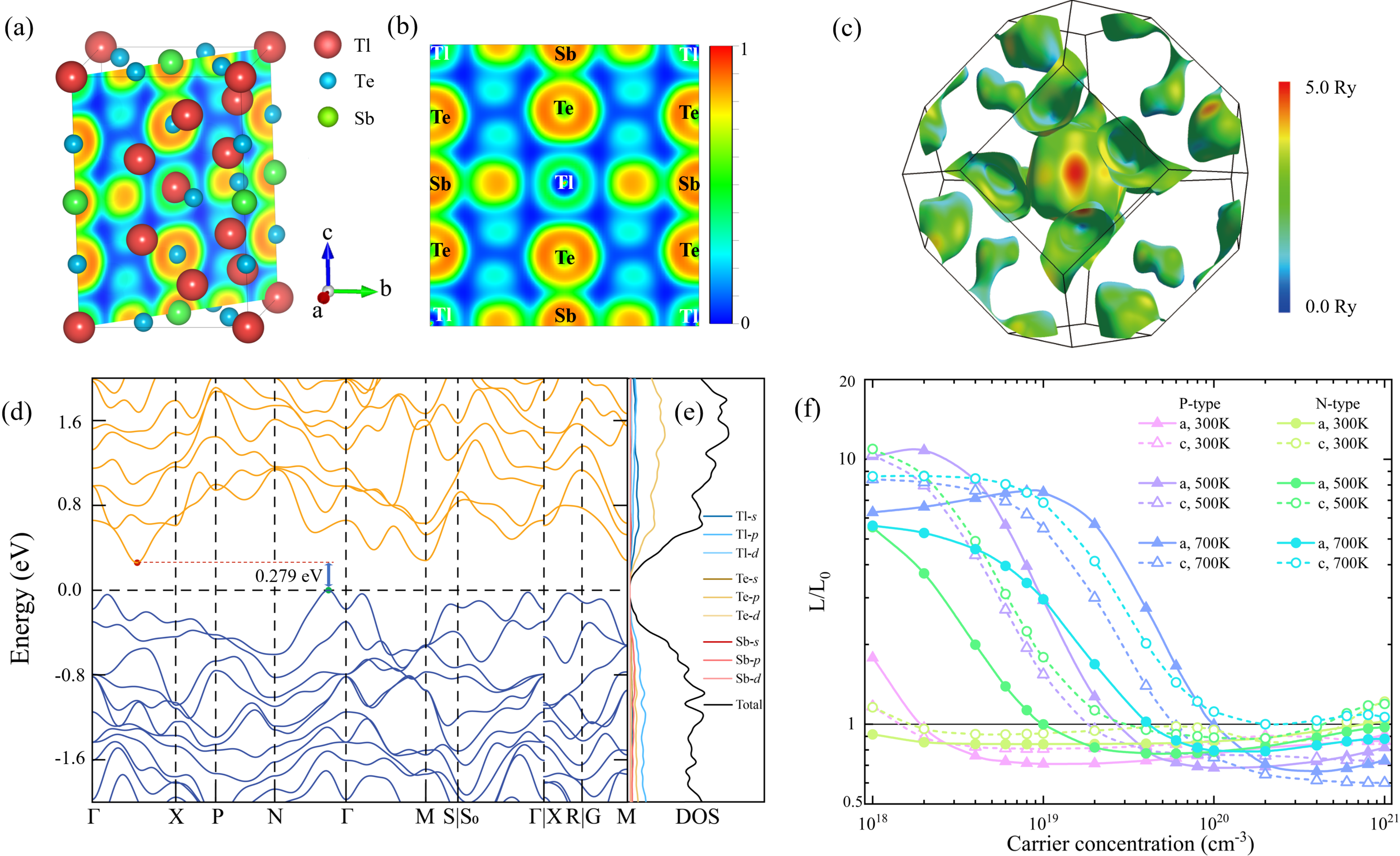}
%\vspace{-2mm}
\caption{Electron Localization Function (ELF) of Tl$_9$SbTe$_6$ in (a) 3D and (b) 2D plots. ELF=1 corresponds to complete localization and ELF=0 corresponds to complete delocalization of electrons. (c) Calculated Fermi surfaces for the valence band structure minimum of Tl$_9$SbTe$_6$ in the first Brillouin zone around the Fermi level. (d) Electronic band structure and (e) electronic density of states (DOS) of Tl$_9$SbTe$_6$. (f) Normalized Lorenz number as a function of carrier concentration at temperatures of 300 K, 500 K and 700 K, respectively. The Lorenz number, denoted by $L_0$, is a constant that is closely approximately by the Sommerfeld value, $L_0 = 2.44\times10^{-8}\ W\ \Omega\ K^{-1}$.
%known as the Lorenz number which is close to the Sommerfeld value $L_0 = 2.44\times10^{-8}\ W\ \Omega\ K^{-1}$.
\label{fig6}}
\end{figure*}
%===========< FIGURE 6 >=========================================

After incorporating $\kappa_c$, the total $\kappa_L$ is approximately 0.27 W m$^{-1}$ K$^{-1}$ and 0.26 W m$^{-1}$ K$^{-1}$ at 400 K and 500 K, respectively. This is consistent with experimental measurements, where $\kappa_L$ is approximately 0.34 and 0.30 W m$^{-1}$ K$^{-1}$ at 400 K and 500 K, respectively~\cite{Guo2014Thermoelectric}. In previous experiment, the $\kappa_L$ is the the difference between total thermal conductivity ($\kappa$) and electronic thermal conductivity ($\kappa_e$), where $\kappa_e$ is estimated via the Wiedemann Franz relationship $\kappa_e = L \sigma T$. The Lorenz number $L$ was calculated by utilizing the single parabolic band and elastic carrier scattering assumption~\cite{Guo2014Thermoelectric}. This implies that the little difference between our $\kappa_L$ and the experimental predicted ones is negligible. Additionally, we observed that the dependence of $\kappa$ on temperature approximately follows $T^{-1/3}$, which indicates the glass-like behaviour of crystalline Tl$_9$SbTe$_6$.

To gain a deeper insight into the thermal transport in complex crystalline Tl$_9$SbTe$_6$, we calculate the spectral and cumulative $\kappa_L$ for both populations' and coherences' contributions, as illustrated in Figs.~\ref{fig3}(a) and (b). Fig.~\ref{fig3}(a) shows that the majority of $\kappa_p$ in Tl$_9$SbTe$_6$ is carried by phonons with a frequency below 2 THz at 300 K. Phonons with frequencies below 0.8 THz at 300 K contribute 57\% and 62\% of $\kappa_p$ in the $a$-axis and $c$-axis directions, respectively. In contrast, as depicted in Fig.~\ref{fig3}(b), the majority of $\kappa_c$ from the wave-like tunneling channel is carried by phonons with a frequency less than 2.4 THz at 300 K. However, phonons with frequencies less than 0.8 THz contribute only 7\% of $\kappa_c$, due to the good particle-like nature of these low-frequency phonons~\cite{He2018High}. Overall, the data presented in Fig.~\ref{fig3}(a) and (b) indicate that low-frequency phonons predominantly contribute to $\kappa_p$, while high-frequency phonons primarily contribute to $\kappa_c$~\cite{Zheng2024Unravelling,wang2023role}. This observation can be attribted to the fact that phonon scattering rates increase with increasing temperature in Tl$_9$SbTe$_6$, as illustrated in Fig.~\ref{fig4}(c). 

%Additionally, we observe a dip in $\kappa_p$ and a peak in $\kappa_c$ around 0.85 THz. This can be attributed to the low-lying flattened modes dominated by Tl atoms with strong scattering. Due to the strong anharmonicity of these modes, they significantly contribute to the coherence conductivity at this frequency range. Figs.~\ref{fig3}(c) and (d) show the contributions to $\kappa_c$ in Tl$_9$SbTe$_6$, calculated using the SCPH+3,4ph model at 300 K and 700 K, resolved in terms of pair phonon energies. From Figs.~\ref{fig3}(c-e), it's evident that quasi-degenerate phonons dominate $\kappa_c$, with optical phonons (frequencies larger than 0.8 THz) contributing the most. Additionally, we observe that the contribution of the couplings between phonons with large frequency differences to $\kappa_c$ in Tl$_9$SbTe$_6$ increases as the temperature rises from 300 K to 700 K in both the $a$ and $c$ directions. This trend is attributed to the increase in phonon scattering rates with temperature.

Furthermore, we observe a dip in $\kappa_p$ and a peak in $\kappa_c$ around 0.85 THz. This can be attributed to the low-lying flattened modes, which are dominated by Tl atoms with strong phonon-phonon scattering. Due to the strong anharmonicity of these modes, they significantly contribute to the coherence conductivity at this frequency range. Figs.~\ref{fig3}(c) and (d) illustrate the contributions to $\kappa_c$ in Tl$_9$SbTe$_6$, calculated using the SCPH+3,4ph model at 300 K and 700 K, respectively, and resolved in terms of pair phonon energies. From Figs.~\ref{fig3}(c-e), it is evident that quasi-degenerate phonons dominate $\kappa_c$, with optical phonons (frequencies larger than 0.8 THz) contributing the most. Additionally, it can be observed that the contribution of the couplings between phonons with large frequency differences to $\kappa_c$ in Tl$_9$SbTe$_6$ increases as the temperature rises from 300 K to 700 K in both the $a$ and $c$ directions. This phenomenon can be attributed to the observed increase in phonon scattering rates with temperature.

%To further understand the microscopic mechanism of the ultralow $\kappa_L$ in Tl$_9$SbTe$_6$, we study several parameters closely related to $\kappa_L$, including phonon velocities, weighted phonon scattering phase space, and phonon scattering rates [see Figs.~\ref{fig4}(a-c)]. Since $\kappa_L$ is proportional to the square of the phonon group velocity ($v^2$), we calculate frequency-dependent $v^2$ where $v$ is the group velocity at 300 K, as shown in Fig.~\ref{fig4}(a). It can be seen that the $v^2$ of Tl$_9$SbTe$_6$ is extremely low, with a maximum value of 5 km$^2\ $s$^{-2}$. In all directions, the maximum value of $v^2$ is lower than that of PbTe, which is around 14 km$^2\ $s$^{-2}$~\cite{Tian2012Phonon}, with $\kappa_L$ of 2 W m$^{-1}$ K$^{-1}$ at 300 K. Also, for the ultralow thermal conductivity material Bi$_4$O$_4$SeCl$_2$ which $\kappa$ is 0.1 W m$^{-1}$ K$^{-1}$ at 300 K, the highest $v$ is around 8 km$^1\ $s$^{-1}$~\cite{Tong2023glass}, which is higher than the maximum $v$ of Tl$_9$SbTe$_6$. Therefore, we mainly attribute the ultralow thermal conductivity in Tl$_9$SbTe$_6$ to the low group velocities. 

To gain further insight into microscopic mechanism of the ultralow $\kappa_L$ in Tl$_9$SbTe$_6$, we conducted a detailed study of several parameters closely related to $\kappa_L$, including phonon velocities, weighted phonon scattering phase space, and phonon scattering rates. These findings are presented in Figs.~\ref{fig4}(a-c). Since $\kappa_L$ is proportional to the square of the phonon group velocity ($v^2$), we calculate frequency-dependent $v^2$ where $v$ is the group velocity at 300 K, as shown in Fig.~\ref{fig4}(a). It can be seen that the $v^2$ of Tl$_9$SbTe$_6$ is extremely low, with a maximum value of 5 km$^2\ $s$^{-2}$. In all directions, the maximum value of $v^2$ is lower than that of PbTe, which is approximately 14 km$^2\ $s$^{-2}$~\cite{Tian2012Phonon}, with $\kappa_L$ of 2 W m$^{-1}$ K$^{-1}$ at 300 K. Furthermore, for the ultralow $\kappa_L$ Bi$_4$O$_4$SeCl$_2$, with a thermal conductivity of 0.1 W m$^{-1}$ K$^{-1}$ at 300 K, exhibits the highest velocity of 8 km$^1\ $s$^{-1}$~\cite{Tong2023glass}, which is higher than the maximum velocity of Tl$_9$SbTe$_6$. Consequently, the ultralow $\kappa$ in Tl$_9$SbTe$_6$ is attributed primarily to the low phonon group velocities. %

%Additionally, we observe the strong scattering rates in crystalline Tl$_9$SbTe$_6$ , as illustrated in Fig.~\ref{fig4}(b)]. More specifically, Fig.~\ref{fig4}(b) clearly shows a strong peak in phonon scattering rates, originating from the low-lying flattening modes dominated by Tl atoms [see Fig.~\ref{fig2}(a)]. In general, the low-lying flattening modes contribute to strong phonon scattering rates, particularly four-phonon scattering rates, thereby suppressing thermal transport~\cite{Li2015Ultralow,Zheng2022Effects}. Therefore, the strong phonon scattering rates arising from the loose bonding of Tl atoms are another factor contributing to the ultralow thermal conductivity in Tl$_9$SbTe$_6$. 

Furthermore, we observe the pronounced scattering rates in crystalline Tl$_9$SbTe$_6$, as illustrated in Fig.~\ref{fig4}(b). In particular, Fig.~\ref{fig4}(b) clearly shows a strong peak in phonon scattering rates, originating from the low-lying flattening modes dominated by Tl atoms [see Fig.~\ref{fig2}(a)]. In general, the low-lying flattening modes contribute to strong phonon-phonon scattering rates, particularly four-phonon scattering rates, thereby suppressing thermal transport~\cite{Li2015Ultralow,Zheng2022Effects,wang2024revisiting,wang2024anomalous}. Therefore, the strong phonon scattering rates arising from the loose bonding of Tl atoms are another factor contributing to the ultralow $\kappa$ in Tl$_9$SbTe$_6$.  

%To gain a better understanding of phonon scattering rates, we also calaulte the phonon scattering phase space, as depicted in Fig.~\ref{fig4}(c). In Fig.~\ref{fig4}(c), the weighted phase space for both three-phonon (3ph) and four-phonon (4ph) scattering increases as the temperature rises from 300 K to 700 K. Obviously, the phase space of 4ph processes is larger than that of 3ph process, resulting in a main source of 4ph scattering rates in Tl$_9$SbTe$_6$. As expected, the peak region around 0.75 THz of the 4ph phase space in Tl$_9$SbTe$_6$ is in coincide with the PDOS of Tl atoms Fig.~\ref{fig2}(b). As mentioned earlier, the low-lying flattening modes generally result in strong four-phonon (4ph) scattering rates by enhancing the 4ph scattering channels. Finally, this will further suppress the thermal transport in Tl$_9$SbTe$_6$.

To obtain a more comprehensive understanding of phonon scattering rates, we also calculated the phonon scattering phase space, as depicted in Fig.~\ref{fig4}(c). As illustrated in Fig.~\ref{fig4}(c), the weighted phase space for both three-phonon (3ph) and four-phonon (4ph) scattering processes increases as the temperature rises from 300 K to 700 K. It is evident that the phase space of 4ph processes is larger than that of 3ph processes, resulting in a significant source of 4ph scattering rates in Tl$_9$SbTe$_6$. As anticipated, the peak region around 0.75 THz of the 4ph phase space in Tl$_9$SbTe$_6$ is in alignment with the partial density of states (PDOS) of Tl atoms, as illustrated in Fig.~\ref{fig2}(b). As previously discussed, the low-lying flattening modes typically result in elevated 4ph scattering rates by enhancing the 4ph scattering channels. Consequently, this will further suppress the thermal transport in Tl$_9$SbTe$_6$. 

\subsection{2.3 Electronic transport properties}
%\textbf{2.3 Electronic transport properties}

%
The calculated electronic transport coefficients including the Seebeck coefficient $S$, electrical conductivity $\sigma$, electronic thermal conductivity $\kappa_e$, and the power factor ($PF=S^2\sigma$) of Tl$_9$SbTe$_6$ based on the Onsager coefficients are presented shown in Fig.~\ref{fig5}. For non-polar crystals, acoustic deformation potential scattering is the main contributor to electric conductivity~\cite{bardeen1950deformation}. Furthermore, we also consider the effects of polar optical phonon scattering~\cite{frohlich1954electrons} and ionized impurity scattering~\cite{AMSET2021}.%

%The calculations of electronic transport performance rely fundamentally on the local density approximation (LDA) method~\cite{Baxter2009Nanoscale}.
%In Fig. 3, we obtain the electrical transport coefficients such as the Seebeck coefficient, electrical conductivity, electronic thermal conductivity, and power factor of Tl$_9$SbTe$_6$, using the momentum relaxation time approximation (MRTA) of the Boltzmann transport equation (BTE), taking into account ionised impurity scattering, acoustic deformation potential scattering, and polar optical phonon scattering which are computed by Ab initio Scattering and Transport (AMSET)~\cite{AMSET2021}.

Fig.~\ref{fig5} presents a depiction of these parameters as functions of carrier concentration at distinct temperatures of 300 K, 500 K, and 700 K, respectively. Notably, Fig.~\ref{fig5}(a) reveals that both p- and n-type doped Tl$_9$SbTe$_6$ manifest notably elevated Seebeck coefficients $S$ across all investigated temperatures, auguring well for the attainment of enhanced $ZT$ values. For instance, at a temperature of 300 K, the maximal Seebeck coefficients for p- and n-doped Tl$_9$SbTe$_6$ materialize at 528 and 393 $\mu$V/K respectively. %Our Tl$_9$SbTe$_6$ has the same order Seebeck coefficient as SnSe which has a favorable 530 $\mu$V/K at room temperature~\cite{zhao2014ultralow}. 
Our Tl$_9$SbTe$_6$ has a Seebeck coefficient that is of a similar order to that of SnSe, which has a favorable 530 $\mu$V/K at room temperature~\cite{zhao2014ultralow}.

%According to the Mott formula\cite{GDMahan1996Mottformula}, the Seebeck coefficient, a parameter pivotal to elucidating the interplay between thermal and electrical characteristics in three-dimensional materials, is expressed as : $S=\frac{2\pi^2k_B^2T}{3q\hbar^2}(\frac{1}{3\pi^2n})^{\frac{2}{3}}m_{DOS,F}^*$, where 'n' denotes carrier concentration and 'm' signifies the effective mass of electron states proximate to the Fermi level. In most cases, when analysing the thermoelectric properties of a material, the density of states effective mass is a parameter related to the isotropic parabolic band effective mass, which differs by a factor of  $N_v^{\frac{2}{3}}$: $m_{DOS}^*=N_v^{\frac{2}{3}}m_{band}^*$, in which $N_v$ is the valley degeneracy~\cite{Nature2011valleydegeneracy}. The valley degeneracy of the Tl$_9$SbTe$_6$ valence band is two, thereby affording a substantial effective mass of density of states and, by extension, superior thermoelectric performance.

The Mott formula~\cite{GDMahan1996Mottformula} indicates that the Seebeck coefficient, a crucial parameter in elucidating the interplay between thermal and electrical characteristics in three-dimensional (3D) materials, 
%, the Seebeck coefficient, a parameter pivotal to elucidating the interplay between thermal and electrical characteristics in 3D materials, 
is expressed as, $S=\frac{2\pi^2k_B^2T}{3q\hbar^2}(\frac{1}{3\pi^2n})^{\frac{2}{3}}m_{DOS,F}^*$, where ``$n$'' and ``$m^*$'' denote the carrier concentration and the the effective mass of electron states around the Fermi level. Furthermore, the density of state effective mass is also a function of band effective mass. The latter is assumed to be a rigid band model, and its dispersion relationship is a standard parabola. They differ by a factor of  $N_v^{\frac{2}{3}}$: $m_{DOS}^*=N_v^{\frac{2}{3}}m_{band}^*$, in which $N_v$ is the valley degeneracy~\cite{Nature2011valleydegeneracy}. The valley degeneracy of the Tl$_9$SbTe$_6$ valence band minimum is two. One is at the $M$ point and the other one is between the $\Gamma$ and $X$ points, as shown in Fig.~\ref{fig6}. Therefore, a favorable $m_{DOS}^*$ will result in a potential high thermoelectric performance for the Tl$_9$SbTe$_6$ material.
%

%Evidently, it can be gleaned from the Mott formula that the Seebeck coefficient for both p- and n-type doping exhibits temperature and doping concentration dependence. As depicted in  Fig.~\ref{fig5}(a), at lower carrier concentrations, the value of S at 300 K reaches the highest. However, as carrier concentration escalates beyond a certain threshold, a sequential decline in the Seebeck coefficient from 700 K to 300 K ensues, aligning with anticipated result. Simultaneously, the data at 300 K exhibits a negative correlation with carrier concentration, demonstrating a monotonic decrement. Notably, at temperatures of 500 K and 700 K, the Seebeck coefficients exhibit peaks, and only when the doping concentration surpasses the peak concentration, does the trend of the Seebeck coefficient variation with the doping concentration shows a regular pattern, whereby an increase in doping concentration leads to a decrease in the value of S. Conversely, when the doping concentration falls below the peak concentration, typically ranging between $10^{19}$ and $10^{20}$ $cm^{-3}$, an anomalous phenomenon emerges. (The elucidation of the underlying reasons behind this anomalous behavior need further explanation.) Furthermore, it is noteworthy that the Seebeck coefficient of Tl$_9$SbTe$_6$ evinces significant anisotropy, a characteristic that remains under equivalent conditions regardless of whether the material for p- or n-type doping.

Moreover, it can be gleaned from the Mott formula that the Seebeck coefficient for both \textit{p}- and \textit{n}-type exhibits a dependence on temperature and doping concentration. As depicted in Fig.~\ref{fig5}(a), at lower carrier concentrations, the value of $S$ at 300 K reaches its highest point. However, as the carrier concentration increases beyond a certain threshold, a sequential decline in the Seebeck coefficient from 300 K to 700 K is observed, aligning with the anticipated result. 

This nonlinear relationship between $S$ and doping concentration stems from the bipolar effect~\cite{zhu2021observation,gong2016investigation}. When there are significant number of both electrons and holes contributing to charge transport (bipolar charge transport) the thermoelectric properties are greatly affected. This occurs when electrons are excited across the band gap producing minority charge carriers (e.g., holes in an n-type material) in addition to majority charge carriers (e.g., the electrons in an n-type material). Bipolar effects are observed in small band gap materials at high temperatures. Furthermore, it is notable that the Seebeck coefficient of Tl$_9$SbTe$_6$ exhibits significant anisotropy, regardless of whether the material is doped with \textit{p}- or \textit{n}-type ions. 
%

%In Fig.~\ref{fig3}(b, c), the computational results for Tl$_9$SbTe$_6$, including both p- and n-type doping scenarios, reveal a conspicuous inverse correlation between electrical conductivity and electronic thermal conductivity with respect to temperature. Simultaneously, there is a direct proportionality observed between these parameters and carrier concentration. Additionally, Tl$_9$SbTe$_6$ demonstrates pronounced anisotropy in both electrical conductivity and electronic thermal conductivity, with distinct behaviors observed for p- and n-type doping scenarios. In the context of electron doping, it is discerned that the values along the c-axis are notably lower than those along the a and b axes. For instance, at a temperature of 300 K and a carrier concentration of 10$^{20}$ cm$^{-3}$, the electrical conductivity and electronic thermal conductivity along the a and b axes measure at 21529.98 S cm$^{-1}$ and 16.32 W m$^{-1}$ K$^{-1}$, approximately twice the values observed along the c-axis (9497.69 S cm$^{-1}$ and 8.46 W m$^{-1}$ K$^{-1}$). Conversely, under hole doping conditions, the values along the c-axis are slightly higher. Moreover, due to the electron's reduced effective mass compared to holes, under identical temperature and carrier concentration conditions, both the electrical conductivity and electronic thermal conductivity for electron-doped Tl$_9$SbTe$_6$ are notably higher along the a and b axes, while displaying the opposite trend along the c-axis. (Reason to be explained here)

Fig.~\ref{fig5}(b) and (e) show the direct correlation between electrical conductivity $\sigma$ and electronic thermal conductivity $\kappa_e$, as well as carrier concentrations and temperatures both \textit{p}- and \textit{n}-type doping of Tl$_9$SbTe$_6$. It can be observed that as temperature increases, conductivity decreases.
%The higher the temperature, the lower the conductivity. 
Conversely, as concentration increases, conductivity increases. 
Furthermore, Tl$_9$SbTe$_6$ exhibits pronounced anisotropy in both electrical conductivity and electronic thermal conductivity, with distinct behaviors observed for \textit{p}- and \textit{n}-type doping scenarios. 
In the context of electron doping, it is evident that the values along the $c$-axis (dashed lines) are notably lower than those along the $a$-axis (solid lines). 
For example, at a temperature of 300 K and a carrier concentration of 10$^{20}$ cm$^{-3}$, the values of $\sigma$ and $\kappa_e$ along the $a$-axis are 6640.35 S cm$^{-1}$ and 4.18 W m$^{-1}$ K$^{-1}$, which are approximately three times the values observed along the $c$-axis with 2254.63 S cm$^{-1}$ and 1.57 W m$^{-1}$ K$^{-1}$. Conversely, under hole doping conditions, the values along the $c$-axis are slightly higher than the corresponding $a$-axis. 

Fig.~\ref{fig5}(c) and (d) exhibit the electronic relaxation time ($\tau_e$) for distinct scattering mechanisms, including acoustic deformation potential (ADP), ionized impurity (IMP) scattering, and polar optical phonon (POP) scattering, respectively. 
It can be seen that $\tau_e$ of IMP is greater than that of ADP, and $\tau_e$ of ADP is greater than POP, indicating that POP has the strongest scattering. As the temperature increases, the strength of electron scattering increases, resulting in a corresponding decrease in $\tau_e$. This is illustrated in Fig.~\ref{fig5}(d), where the total electronic relaxation time is observed to be on the order of 10$^{-14}$ cm$^-3$. 
%electron scattering becomes stronger and stronger, causing the corresponding $\tau_e$) to become smaller and smaller, as shown in Fig.~\ref{fig5}(d). The total electronic relaxation is basically on the order of 10$^{-14}$ cm$^-3$. 
As the temperature increases from 300 K to 700 K, the electron relaxation time decreases from 2.4 $\times$ 10$^{-14}$ cm$^-3$ to 1.0 $\times$ 10$^{-14}$ cm$^-3$ for \textit{p}- and \textit{n}-type doping Tl$_9$SbTe$_6$. %

%As illustrated in Fig.~\ref{fig3}(d), based on the Seebeck coefficient and electrical conductivity, we have computed the power factor $(S^2\sigma)$under varying temperatures and carrier concentrations. Our findings reveal that Tl$_9$SbTe$_6$ exhibits substantial anisotropy in its power factor. Across diverse temperature and carrier concentration regimes, the power factor for electron-doped Tl$_9$SbTe$_6$ along the a and b axes consistently surpasses that for hole doping. This underscores that, to enhance the thermoelectric performance of Tl$_9$SbTe$_6$, electron doping is more advantageous than hole doping. In contrast, the effect of hole doping along the c-axis is notably more pronounced and superior in enhancing thermoelectric performance. The variation of power factor with concentration and temperature is not monotonic as a function of electrical and electronic thermal conductivity. It exhibits a characteristic trend whereby the power factor initially rises with increasing carrier concentration and subsequently declines, with a peak concentration ranges between 10$^{20}$ and 10$^{21}$ cm$^{-3}$. Below this peak concentration, higher temperatures are associated with lower power factors. However, beyond the peak concentration, the power factor exhibits an upward tendency with increasing temperature. Notably, at a temperature of 500 K and a carrier concentration of 2×10$^{20}$ cm$^{-3}$, the power factor along the c-axis attains its maximum value, reaching 8.57 (mW m$^{-1}$ K$^{-2}$).

As illustrated in Fig.~\ref{fig5}(f), the power factor $(S^2\sigma)$ was computed based on the Seebeck coefficient and electrical conductivity, under varying temperatures and carrier concentrations. Our findings reveal that Tl$_9$SbTe$_6$ exhibits substantial anisotropy in its power factor. Across intermediate temperature and carrier concentration regimes, the power factor for electron-doped Tl$_9$SbTe$_6$ along the $a$-axis consistently surpasses that for hole doping. This indicates that, in order to enhance the thermoelectric performance of Tl$_9$SbTe$_6$, electron doping is more advantageous than hole doping. In contrast, the effect of hole doping along the $c$-axis is notably more pronounced and superior in enhancing the power factor. The variation of power factor with concentration and temperature is not monotonic as a function of electrical and electronic thermal conductivity. The power factor exhibits a characteristic trend whereby it initially rises with increasing carrier concentration and subsequently declines, with a peak concentration range between 10$^{20}$ and 10$^{21}$ cm$^{-3}$ for Tl$_9$SbTe$_6$. Below this peak concentration, higher temperatures are associated with lower power factors. It is noteworthy that at a temperature of 500 K and a carrier concentration of 2×10$^{20}$ cm$^{-3}$, the power factor along the $c$-axis attains its maximum value, reaching 8.57 mW m$^{-1}$ K$^{-2}$. 
%

%Fig.~\ref{fig4}(c) meticulously presents the electronic band structure and the total and partial DOSs. Tl$_9$SbTe$_6$ emerges as an indirect bandgap semiconductor, characterized by its conduction band minimum (CBM) positioned between the $\Gamma$ and X points, and its valence band maximum (VBM) located between the N and $\Gamma$ points, resulting in a discernible bandgap of 0.279 eV. The contributions to the bottom of the conduction band and the top of valence band predominantly arise from the p-orbitals of Te atoms and Tl atoms respectively. Furthermore, the band structure reveals a dual-degeneracy phenomenon within the valence band at the M point. Importantly, this observation holds profound significance for enhancing the thermoelectricity (ZT) of the material, as the insights from band engineering studies~\cite{Nature2011valleydegeneracy,2012bandengeering}. 

Fig.~\ref{fig6}(a) and (b) show the 3D and 2D Electron Localization Function (ELF) of Tl$_9$SbTe$_6$. ELF values of 1 and 0 represent complete localization and delocalization of electrons. 
%ELF=1 corresponds to complete localization and ELF=0 corresponds to complete delocalization of electrons. 
The Te and Sb elements exhibit a stronger tendency to attract electrons than the Tl elements, suggesting a greater degree of delocalization for thallium atoms. The delocalized electrons show a stronger electron-phonon coupling~\cite{gao2021highly,qin2022switch}.  
%attract more electrons than Tl elements, indicating that a more delocalization of electrons for thallium atoms. These delocalized electrons scatter more phonons, which means stronger electron-phonon coupling~\cite{gao2021highly}. 
This is evident in Fig.~\ref{fig2}(b), where the thallium element exhibits a very pronounced peak in the low-frequency region (\textless 1.0 Thz) for the phonon density of states~\cite{gao2020thermoelectric,gao2018high}.

Fig.~\ref{fig6}(d) and (e) present the electronic band structure and the electronic density of state (DOS). Tl$_9$SbTe$_6$ emerges as an indirect bandgap semiconductor, characterized by its conduction band minimum (CBM) positioned between the $\Gamma$ and $X$ points, and its valence band maximum (VBM) located between the $N$ and $\Gamma$ points, resulting in a discernible bandgap of 0.279 eV. The contributions to the bottom of the conduction band and the top of the valence band predominantly arise from the $p$-orbitals of Te atoms and Tl atoms respectively. Furthermore, the band structure reveals a dual-degeneracy phenomenon within the valence band at the $M$ point. Importantly, this observation holds profound significance for enhancing the thermoelectricity $ZT$ of the material, as the insights from band engineering studies~\cite{Nature2011valleydegeneracy,2012bandengeering}. 

It is well established that electrical transport is not solely dependent on the movement of electrons, but also on the transfer of heat. This phenomenon is commonly described by the Wiedemann-Franz (WF) law, which states that the ratio of electrical conductivity to the conductivity of heat ($\kappa_e$/$\sigma$) is directly proportional to the absolute temperature $T$. This relationship can be expressed as $ \kappa_e/ \sigma = LT$, where $L$ is a constant ratio known as the Lorenz number. The value of $L$ is believed to be close to the Sommerfeld value $L_0$, which is approximately \[2.44\times10^{-8}\ W\ \Omega\ K^{-1}\]~\cite{Sangwook2017Anomalously,Wen2023Wiedemann}. It is notable that most metals that considered to be ``good'' in terms of their conductivity satisfy this WF law. However, the Lorenz number of transition metals such as palladium (Pd), nickel (Ni), cobalt (Co), and platinum (Pt) exhibits larger deviations from the Sommerfeld value $L_0$~\cite{tong2019comprehensive,zhang2022pressure}. Recently, a $\kappa_e$ that is an order of magnitude lower than expected from WF law was observed in metallic VO$_2$ near its insulator-metal transition. This phenomenon was explained in terms of the absence of quasiparticles in a strongly correlated electron fluid in which heat and charge diffuse independently~\cite{Sangwook2017Anomalously}.

The calculated doping-dependent ratio is shown in Fig.~\ref{fig6}(f). It can be seen that when the doping concentration is high, especially when it exceeds 10$^{20}$ cm$^{-3}$, $L$ is near equivalent to $L_0$, exhibiting metallic properties. The lower limit of the doping concentration for the Lorenz number being close to $L_0$ increases with temperature. For \textit{p}-type Tl$_9$SbTe$_6$, the lower limit of the doping concentration for the Lorenz number to be close to $L_0$ at 300 K is around 10$^{18}$ cm$^{-3}$, while at 700 K it is 6$\times$10$^{19}$ cm$^{-3}$. This suggests that in order to achieve enhanced thermoelectric performance under high temperature conditions, it may be necessary to adjust the doping concentration within an optimal range. 
%Also the L value of n-type in the $c$ direction is close to that of p-type, while the L values in $a$ and $b$ directions are closer to $L_0$ than that of p-type.

A higher $L$ value represents a higher ratio of electronic thermal conductivity to electrical conductivity, which is less conducive to improving the thermoelectric properties of the material. At lower concentrations, a substantial deviation of $L$ from the ideal value $L_0$ is observed, particularly when the concentration is below 5$\times$10$^{19}$ cm$^{-3}$. Furthermore, the deviation of $L$ from $L_0$ is severer with higher temperature than lower temperature. This is consistent with the trend shown in Fig.~\ref{fig7}, where the lower the doping concentration, the closer the $ZT$ peak value is to the low temperature region. This also makes it difficult to use the Wiedemann-Franz law to predict the $\kappa_L$ of the material by considering the experimental values of $\kappa$ and electrical conductivity, as considering $L$ as $L_0$ will underestimate $\kappa_e$, thus overestimate $\kappa_L$. Consequently, is is necessary to calculate the Lorenz number $L$ as a function of temperature and doping concentration, which is the prerequisite for accurately estimating thermoelectric performance.

\subsection{2.4 Thermoelectric performance}
%\textbf{2.4 Thermoelectric performance}

%===========< FIGURE 7 >=========================================
\begin{figure*}%[htp]
\includegraphics[width=1.75\columnwidth]{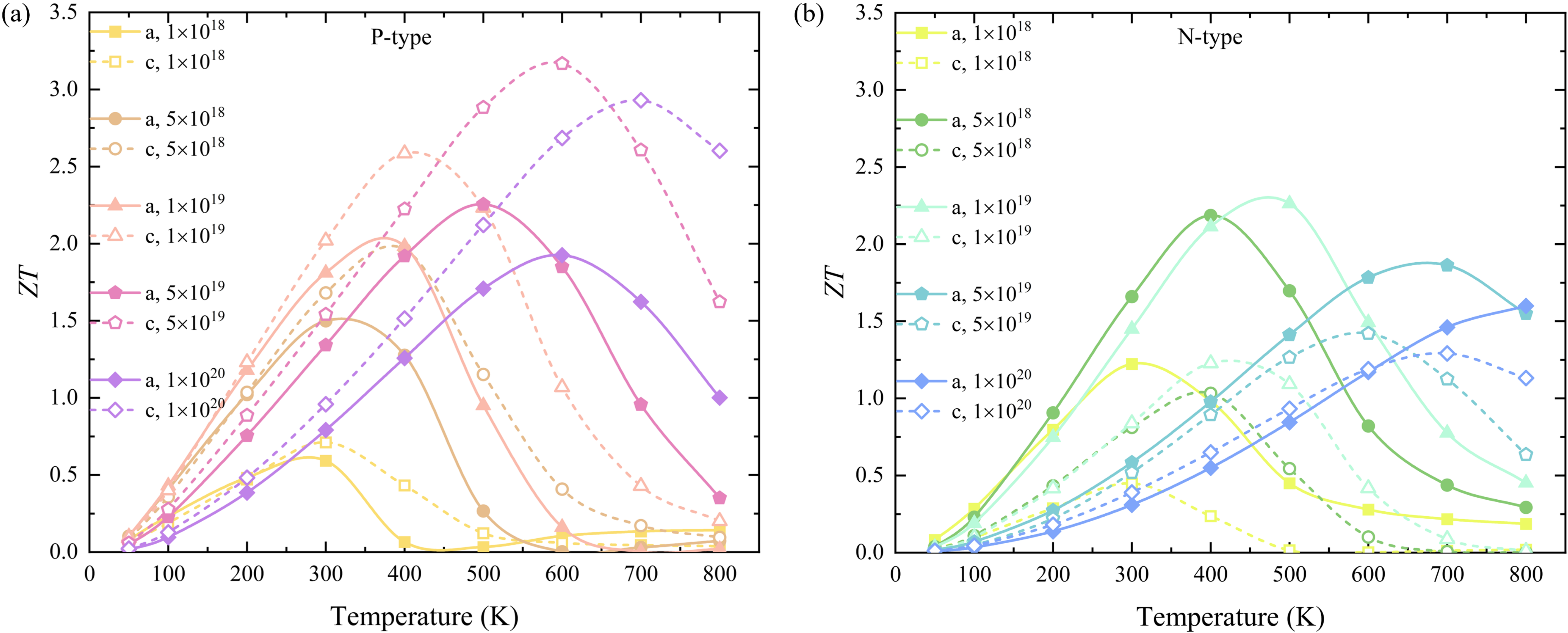}
%\vspace{-2mm}
\caption{The calculated figure of merit ($ZT$) of (a) \textit{p}-type and (b) \textit{n}-type Tl$_9$SbTe$_6$ as a function of temperature along the $a-$ and $c-$ axes at different carrier concentrations, ranging from 1.0$\times$10$^{18}$ cm$^{-3}$ to 1.0$\times$10$^{20}$ cm$^{-3}$.
\label{fig7}}
\end{figure*}
%===========< FIGURE 7 >=========================================

The calculated $ZT$ of Tl$_9$SbTe$_6$ at different temperatures and doping concentrations is presented in Fig.~\ref{fig7}. As the doping concentration increases from 1.0$\times$10$^{18}$ cm$^{-3}$ to 1.0$\times$10$^{20}$ cm$^{-3}$, the peak value of the $ZT$ value will vary from 300 K to 700 K and higher, with a greater bias towards higher temperatures. Concurrently, the peak value will initially increase and subsequently decrease, reaching a maximum value at a doping concentration between 5$\times$10$^{18}$ cm$^{-3}$ and 1$\times$10$^{19}$ cm$^{-3}$. Due to the orientation-dependent properties of phonons and electrons, the $ZT$ value will exhibit a significant anisotropy. 

In the case of \textit{p}-type doping, the $ZT$ value in the $c$-axis direction is significantly higher than that in the $a$-axis direction, which is the opposite in the case of \textit{n}-type. At the same temperature, the $ZT$ values of \textit{p}- and \textit{n}-type vary significantly. The highest $ZT$ value of the \textit{p}-type can reach a peak of 3.17 at 600 K, with a doping concentration of 5$\times$10$^{19}$ cm$^{-3}$. The highest $ZT$ value of the \textit{n}-type also reaches a peak of 2.26 at 500 K, with a doping concentration of 1.0$\times$10$^{19}$ cm$^{-3}$. It can be concluded that the Tl$_9$SbTe$_6$ is a highly promising thermoelectric material in the medium temperature zone (300 K-600 K), particularly for the \textit{p}-type doping.

\section{3. Conclusion} 

In conclusion, our study examines the thermal and electrical transport properties of Tl$_9$SbTe$_6$ through the integration of machine-learning-potential-based self-consistent phonon (SCPH) theory, first-principles calculations, and a unified theory of thermal transport that incorporates both coherence and population conductivities. The Moment Tensor Potential, trained with Machine Learning Inter-atomic Potential (MLIP), achieves DFT-level accuracy in predicting energy and atomic forces, thereby accurately producing the phonon dispersions for crystalline Tl$_9$SbTe$_6$. The force constants derived from the MLIP and SCPH theory indicate a phonon hardening phenomenon across the entire frequency range, which suggests significant lattice anharmonicity in Tl$_9$SbTe$_6$. Subsequently, an ultralow $\kappa_L$ for Tl$_9$SbTe$_6$ is predicted, with a value of 0.3 W m$^{-1}$K$^{-1}$ at room temperature. This value is consistent with the experimentally measured $\kappa_L$ and exhibits a glass-like temperature dependence.

%In brief, our research reveals significant impacts of four-phonon interaction processes and off-diagonal elements of heat flux operators on thermal transport in Tl$_9$SbTe$_6$. Specifically, the four-phonon interaction processes influence phonon energy shifts, which increase thermal conductivity, and also significantly enhance overall scattering rates, thereby decreasing thermal conductivity. Additionally, we find that in crystalline Tl$_9$SbTe$_6$ at temperatures above 600 K, the coherent component $\kappa_c$ is the dominant contributor to the lattice $\kappa_L$, with a $\kappa_c/\kappa_p$ ratio larger than 1. By analyzing the electronic band structure, we find that Tl$_9$SbTe$_6$ also exhibits good electrical transport properties, attributed to the dual-degeneracy phenomenon within the valence band. 

Our research indicates that four-phonon interaction processes and off-diagonal elements of heat flux operators have a significant impact on thermal transport in Tl$_9$SbTe$_6$. Specifically, the four-phonon interaction processes influence phonon energy shifts, which increase thermal conductivity. However, they also significantly enhance overall scattering rates, thereby decreasing thermal conductivity. Furthermore, we observe that in crystalline Tl$_9$SbTe$_6$ at temperatures above 600 K, the coherent component $\kappa_c$ represents the dominant contributor to the lattice $\kappa_L$, with a $\kappa_c/\kappa_p$ ratio exceeding 1. By analyzing the electronic band structure, we find that Tl$_9$SbTe$_6$ also exhibits excellent electrical transport properties, attributed to the dual-degeneracy phenomenon within the valence band. 

%Finally, our investigation reveals that Tl$_9$SbTe$_6$, with a maximum figure of merit (ZT) of up to 3.17, emerges as a promising candidate for thermoelectric applications, owing to its ultralow thermal conductivity and favorable electrical transport properties. We also elucidate the origins of the ultralow thermal conductivity and relatively high electrical conductivity in Tl$_9$SbTe$_6$, confirming its optimal charge carrier concentration and operating temperature range. Our study not only offers insights into potential improvements and application directions for the Tl$_5$Te$_3$ family of materials but also paves the way to accelerate the accurate prediction of thermoelectric materials.

Our investigation has revealed that Tl$_9$SbTe$_6$, with a maximum figure of merit (ZT) of up to 3.17, is a promising candidate for thermoelectric applications. This is due to the ultralow $\kappa_L$ and favorable electrical transport properties. We have also elucidated the origins of the ultralow $\kappa_L$ and relatively high electrical conductivity in Tl$_9$SbTe$_6$. This has involved confirming its optimal charge carrier concentration and operating temperature range. Our study offers insights into potential improvements and application directions for the Tl$_5$Te$_3$ family of materials and paves the way to accelerate the accurate prediction of thermoelectric materials.

%================================================================
\section{ACKNOWLEDGMENTS}
We acknowledge the support from the National Natural Science Foundation of China 
(No.12104356 and No.52250191, No.12104236), 
% 12104356 and 52250191 Zhibin Gao;
% 12104236 Wenjie Hou;
the Opening Project of Shanghai Key Laboratory of Special Artificial Microstructure Materials 
and Technology (No.Ammt2022B-1), and the Fundamental Research Funds for the Central 
Universities. 
% Ammt2022B-1 from Zhibin Gao;
H. Gu acknowledges the support from Gusu Leading Talent  (No.ZXL2021383).  % Suzhou Lab
This work is sponsored by the Key Research and Development Program of the Ministry of Science and Technology (No.2023YFB4604100).
% No.2023YFB4604100 from Zhibin Gao;
We also acknowledge the support by HPC Platform, Xi’an Jiaotong University. 
%W. Shi acknowledges the support from the National Natural Science Foundation of China (22103099), the Guangzhou Science and Technology Plan Project (202201011155)

%===============================
%This can yield the finding that in scenarios where the lattice thermal conductivity κL is exceptionally low, the incorporation of high-order anharmonicity and off-diagonal terms serves to reconcile the disparities between experimental observations and theoretical predictions.

 \bibliography{References} %your bib file here
 \end{document}